\documentclass[12pt]{iopart}
\usepackage{graphicx} 
\font\tenimbf=cmmib10 at 10pt
\font\sevenimbf=cmmib10 at 6pt
\font\fiveimbf=cmmib10 at 4pt
\newfam\imbf
\textfont\imbf=\tenimbf
\scriptfont\imbf=\sevenimbf
\scriptscriptfont\imbf=\fiveimbf

\def\bea{\begin{eqnarray}}
\def\eea{\end{eqnarray}}
\def\beq{\begin{equation}}
\def\eeq{\end{equation}}
\begin{document}

\title[Inclusive LHC distributions compared with an adjusted
DPMJET-III model]{Inclusive distributions in p-p collisions at  LHC
energies compared with  an adjusted  DPMJET-III model with chain fusion}

\author{F.Bopp$^1$ and  J.Ranft$^1$  }

\address{$^1$ Siegen University, Siegen, Germany}
    \setlength{\textheight}{23cm}
     \setlength{\textwidth}{15cm}
%
%
         \setlength{\topmargin}{-0cm}
         \setlength{\oddsidemargin}{0pt}

\begin{abstract}
A DPMJET-III model (DPMJET-III-2011) with chain fusion adjusted 
to include energy.dependent parameters 
  is used to calculate inclusive
    distributions in p--p  collisions at LHC energies.
     Presented are charged hadrons  
     rapidity distributions, transverse momentum
     distributions, multiplicity distributions as well as
     multiplicities at mid-rapidity as function of the
    collision energy. For hadrons with strangeness we present
    cms-rapidity distributions and transverse momentum
    distributions. 
With the considered merely energy-dependent adjustments the obtained agreement with
the transversal $\Lambda$ and $\Xi$  distribution is not satisfactory. 
\end{abstract}

\maketitle 

\section{Introduction} 

Monte Carlo codes based on the two--component Dual Parton Model
involving soft and hard  hadronic collisions producing chains of
particles are
available since almost 20 years~{\cite{Aurenche:1994ev}}. 
The present codes are:\vspace{0.2cm}

\hspace{2cm}\begin{tabular}{ll}
PHOJET  &  for h--h and $\gamma$--h collisions \cite{phojet-a}\tabularnewline 
DPMJET-III &  for h--h, h--A and A--A collisions \cite{dpmjet3}\tabularnewline
\end{tabular}\vspace{0.2cm}

In distinction to earlier versions DPMJET-III 
is based on PHOJET for its h--h  collisions. 
In such collision it is therefore - except for a few additions 
like the fusion discussed below - identical to PHOJET. 
PHOJET describes the production of strings. For the 
string decay it calls PYTHIA version
6.412~\cite{Sjostrand:2006za}.
For a few special cases we found it necessary to change the
PYTHIA fragmentation.
These were done in the DPMJET part,
leaving the PYTHIA code itself  untouched.

As we use the full program we will refer below just to DPMJET-III. We now
outline its main additions.
 
Comparing DPMJET-III to RHIC data it was learned that something had to be 
done to decrease the particle density. As the strings  are quite dense in
impact parameter space  interactions between strings are plausible. The
expected percolation was modeled as fusion of close hadronic chains 
implemented  in DPMJET-III~\cite{dpmfusion1} in 2004.  The obtained reduction  
was very essential  for central collisions of
heavy ions, but fusion also  changes the particle production in 
very high energy p--p collisions when the number of  contributing 
chains obtained by a Glauber / eikonal formalism gets sizable.

RHIC and Fermilab data also contain interesting information about particle antiparticle
ratio's~\cite{Bopp:2005cr}. For the baryon/antibaryon distribution the string fusion mentioned 
above can be significant (p.e. two quark-antiquark strings can fuse to 
a diquark-antidiquark string yielding baryons and antibaryons). 

In the  
diquark string decay used in PYTHIA one observes a dip in the ratio of the $\Omega /
\bar{\Omega}$ spectra not seen in the data. A solution of the problem is to
include a small contribution of diquark-diantiquark mesons production in the
first rank so that $\Omega$ can appear in the second rank. The idea is that
such tetra-quark mesons are always produced  but decay too fast to be identified in mass plots.

The LHC experiments did compare DPMJET-III to particle
production at LHC--energies, see \cite{cmsdpmjet},
\cite{alicedpmjet} and \cite{atlasdpmjet}. There were some
successful predictions of DPMJET. However, LHC
experiments found that around 7 TeV the multiplicity rises faster with energy 
than predicted by DPMJET-III. 

In order to make the program usable for ongoing data analyses  
at LHC energies  we  adjusted the program to improve  the agreement with
available experimental results.  
We allowed for an energy-dependence of 
string decay parameters. No differentiation between softer and harder
strings was attempted.
The new results of this modified version   will be reported  in
section 3 and 4.

\section{Modifications of DPMJET-III needed for LHC energies}

There are essentially three additional modifications 
of DPMJET-III implemented 
in order to get better agreement with LHC data on particle production
in p--p   collisions.
\begin{description}

\item [{(1)}] The first modification is connected to a problem with
collision scaling known since 2004~\cite{collsca}. DPMJET-III uses an eikonal formalism
to determine the size of various multiple scattering contributions
$P_{n,\{\alpha_i^f\},\{\alpha_i^b\}}$ where $n$ is the number of chains and
$\alpha_{i}^{f/b}$
is the Regge intercept depending on the diquark, valence quark or sea quark
nature of the forward/backward parton of  $i$-th chain. Let us consider the
forward direction. 
For each such configuration the attributed energy fractions $\{x_i\}$ to
these  partons are   then chosen with a factorizing structure function of the form:
\[
P_{n,\{\alpha_i\}}\int\prod_{i}^{n}x_{i}^{\alpha_{i}-1}\delta(1-\sum_{j}^{n}x_{j})\]
in which the energy available for a scattering process depends on 
the reminder. 
There are two kinds of chains in DPMJET.
Hard chains produced by hard collisions of partons from the colliding
hadrons  (typically
large $p_{\bot}$) and soft chains representing soft hadron production 
in the collisions. In the factorizing formalism 
soft processes affect the energy sampled in hard processes.
\\
This turned out to be too simple. Experiments~\cite{Klay:2004ma} gave
evidence for collision scaling in not too central scattering 
processes\footnote{ For central heavy ion
collision collision scaling is lost as transport effects become important. So far these
effects are not implemented. }.
Collision scaling means, that exactly as many hard chains are produced
as predicted by considering just hard collisions. \\
To correct for the missing collision scaling  an additional parameter was  
introduced~\cite{collsca}  which increases
the number of hard collisions in such a way that collision scaling
is obtained. We here adjusted these constants to the LHC data.

\item [{(2)}] LHC experiments \cite{cmsdpmjet,alicedpmjet,atlasdpmjet}
found that the produced hadron multiplicity rises with energy 
somewhat faster
than the model. To obtain this faster rise with collision energy one
needs to introduce energy-dependent parameters. This
is not unreasonable. The time scale of the initial scattering is inversely
proportional to the energy. This causes a more localized 
string and  a widening of the $p_{\bot}$distribution
of the string ends which was observed as on contribution to  
the multiplicity dependence of the average
transverse momentum  $<p_{\bot}>_{n}$.
There are of course many different parameter in PYTHIA which
can be tuned in an energy-dependent way.\\
The solution which was adopted for the moment is not to search for a
suitably, possibly theoretically plausible energy-dependence but
just adjust parameters. 
The two parameters which determine the multiplicity
of fragmenting chains are the Lund parameters {\em{PARJ(41)}}  and
{\em{PARJ(42)}} for which we use the following values:

\noindent \begin{eqnarray}
PARJ(41)=\nonumber 
 & 0.2 & {~~}\mathrm{ for}{~}{E_{cm}\le3\,\mathrm{TeV}}\nonumber \\
 & 0.2+0.1(E_{cm}-3)/4 & {~~}\mathrm{ for}{~}E_{cm}\in[3\,\mathrm{TeV},7\,\mathrm{TeV}]\nonumber \\
 & 0.3+0.05(E_{cm}-7)/7 & {~~}\mathrm{ for}{~}E_{cm}\in[7\,\mathrm{TeV},14\,\mathrm{TeV}]\nonumber \\
 & 0.35 & {~~}\mathrm{ for}{~}E_{cm}\ge14\,\mathrm{TeV} \label{parj41}
\\
PARJ(42)=\nonumber 
 & 0.8 & {~~}\mathrm{ for}{~}{E_{cm}\le3\,\mathrm{TeV}}\nonumber \\
 & 0.8-0.2(E_{cm}-3)/4 & {~~}\mathrm{ for}{~}E_{cm}\in[3\,\mathrm{TeV},7\,\mathrm{TeV}]\nonumber \\
 & 0.6-0.1(E_{cm}-7)/7 & {~~}\mathrm{ for}{~}E_{cm}\in7\,\mathrm{TeV},14\,\mathrm{TeV}]\nonumber \\
 & 0.5 & {~~}\mathrm{ for}{~}E_{cm}\ge14\,\mathrm{TeV}  \label{parj42} \end{eqnarray}
 In (\ref{parj41}) and (\ref{parj42}) we do not continue to change
 {\em{PARJ(41)}}  and {\em{PARJ(42)}}
for $E_{cm}$ larger than the maximum LHC energy
of 14000 GeV. For the moment (in "DPMJET-III-2011")  we replace in the FORTRAN code PARJ(41)
and PARJ(42) by the values given in (\ref{parj41}) and (\ref{parj42})
as soon as in the input cards a change of PARJ(41) and PARJ(42) is
demanded. 

\item [{(3)}] The third modification is connected to the
production of strange hadrons. The production of  $K_s^0$
mesons and of $\Lambda$ and $\Xi ^-$ hyperons  in p-p collisions was 
measured  by the CMS Collaboration  \cite{CMSstrange}. 
 The program gave more $K_s^0$
 than measured by CMS, while it obtained less
 $\Lambda$ and $\Xi ^-$ production than measured by CMS.
To increase the agreement with the measurements in this regard 
more energy-dependent parameters have to be introduced. 

Hyperon and strange meson
production in DPMJET-III is controlled by the Lund parameters
PARJ(1), PARJ(2), PARJ(3), PARJ(5) and PARJ(6). 
The default of the parameter PARJ(2) was not touched. For the other parameters the
following energy-dependent values  (in the energy
range up to $E_{cm}$ = 14 TeV) were implemented:

\end{description}
\noindent \begin{eqnarray}
PARJ(1)=\nonumber 
 & 0.1 & {~~}\mathrm{ for}{~}{E_{cm}\le 0.5\,\mathrm{TeV}}\nonumber \\
 & 0.1+0.1(E_{cm}-0.5)/0.4 & {~~}\mathrm{for}{~}E_{cm}\in[0.5\,\mathrm{TeV},0.9\,\mathrm{TeV}]\nonumber \\
 & 0.2 & {~~}\mathrm{for}{~}E_{cm}\ge[0.9\,\mathrm{TeV}]\label{parj1}
\\
PARJ(3)=\nonumber 
 & 0.4 & {~~}\mathrm{ for}{~}{E_{cm}\le 0.5\,\mathrm{TeV}}\nonumber \\
 & 0.4+1.6(E_{cm}-0.5)/0.4 & {~~}\mathrm{for}{~}E_{cm}\in[0.5\,\mathrm{TeV},0.9\,\mathrm{TeV}]\nonumber \\
 & 2.0 & {~~}\mathrm{for}{~}E_{cm}\ge[0.9\,\mathrm{TeV}]\label{parj3}
\\
PARJ(5)=\nonumber 
 & 0.5 & {~~}\mathrm{ for}{~}{E_{cm}\le 3.0\,\mathrm{TeV}}\nonumber \\
 & 0.5-0.05(E_{cm}-3.0)/4.0 & {~~}\mathrm{for}{~}E_{cm}\in[3.0\,\mathrm{TeV},7.0\,\mathrm{TeV}]\nonumber \\
 & 0.45 & {~~}\mathrm{for}{~}E_{cm}\ge[7.0\,\mathrm{TeV}]\label{parj5}
\\
PARJ(6)=\nonumber 
 & 0.5 & {~~}\mathrm{ for}{~}{E_{cm}\le 1.0\,\mathrm{TeV}}\nonumber \\
 & 0.5+0.55(E_{cm}-1.0)/6.0 & {~~}\mathrm{for}{~}E_{cm}\in[1.0\,\mathrm{TeV},7.0\,\mathrm{TeV}]\nonumber \\
 & 1.05 & {~~}\mathrm{for}{~}E_{cm}\ge[7.0\,\mathrm{TeV}]\label{parj6}
\\ \nonumber
\end{eqnarray}

\section{Comparison of DPMJET--III--2011 results with LHC Data on
charged hadron production} 

We start to discuss the {\em nsd} (non single diffractive)  
and  {\em inel} (inelastic)  pseudo-rapidity distribution  $dN_{ch}/d\eta_{cm}$ 
measured by the CMS and ALICE Collaborations.

\includegraphics[width=12.5cm]{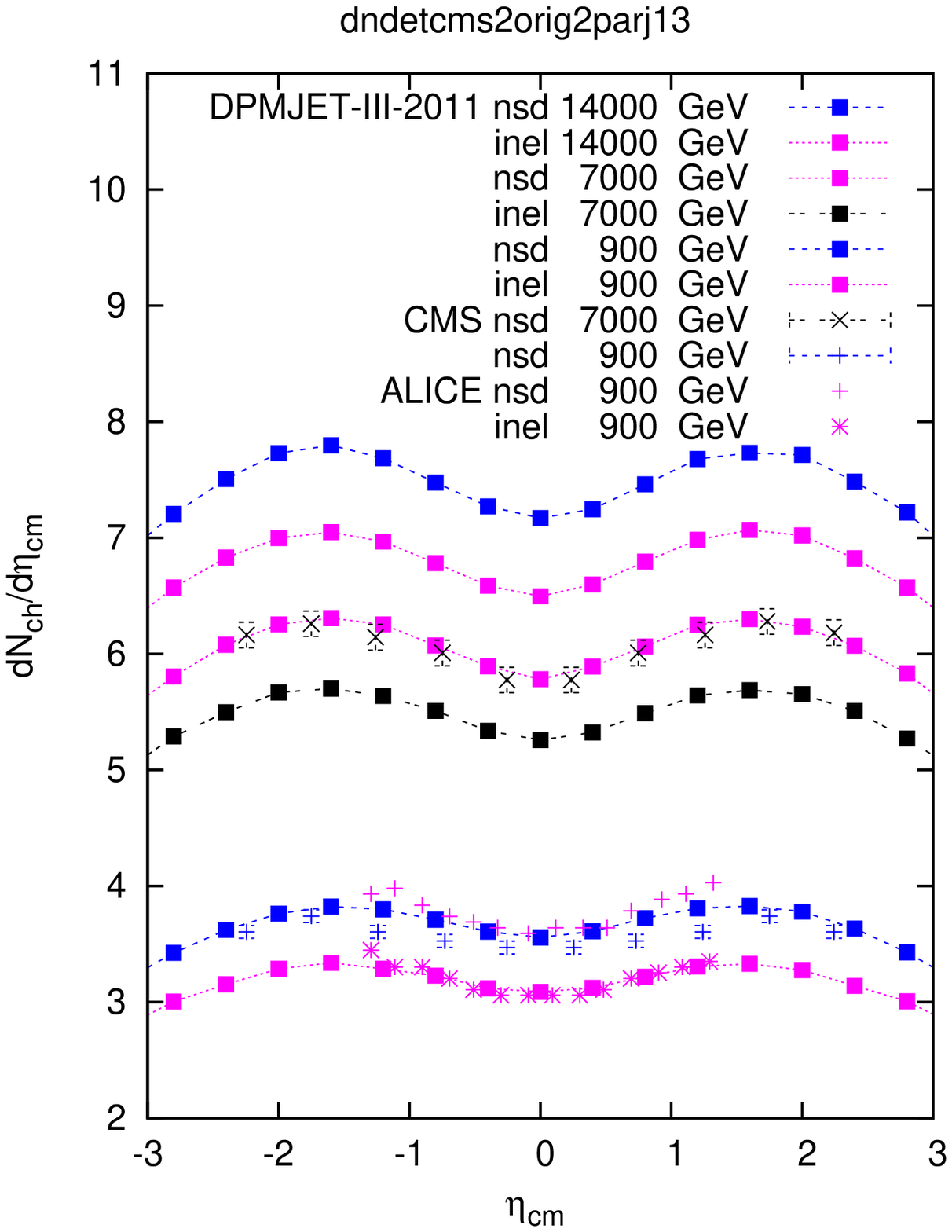}

{{\bf Fig.1} Central $\eta_{cm}$ distributions of charged particles in
$\sqrt (s)$ = 900, 7000 and 14000 GeV p--p collisions compared to 
non single diffractive ({\em nsd}) and  inelastic ({\em inel}) 
pseudo-rapidity distributions
obtained with DPMJET-III-2011. The experimental data 
for  {\em nsd} and  {\em inel} collisions are from the CMS ({\em nsd})  
\cite{cmsnsd} and from the ALICE \cite{alicensdinel} collaboration.}

\newpage
In Fig.1 and Fig.2 we present for p--p collisions the
non single diffractive data from CMS \cite{cmsnsd} at 
900, 2360 and 7000 GeV and
the non single diffractive and inelastic data 
from ALICE \cite{alicensdinel} at 900 and
2360 GeV and compare them with the results from
DPMJET--III-2011. Excellent agreement is obtained.

Also included are the results  at  9000 and
14000 GeV in p--p collisions. At these energies the distributions
are expected to be  be measured at the LHC 
in future. 

\includegraphics[width=13.5cm]{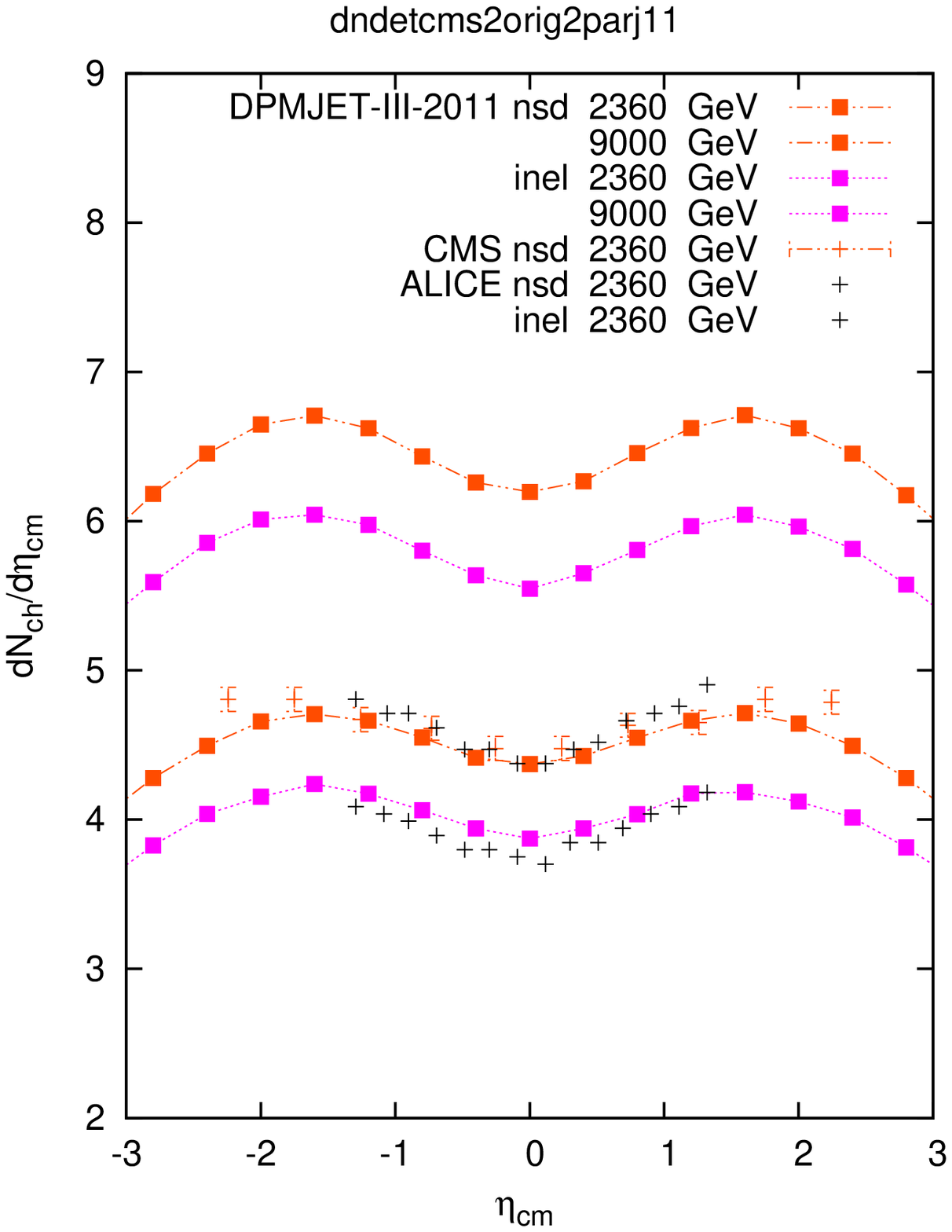}

{{\bf Fig.2} Additional central $\eta_{cm}$ distributions 
of charged particles in
$\sqrt (s)$ = 2360 and 9000 GeV p--p collisions compared to 
non single diffractive ({\em nsd}) and  inelastic ({\em inel}) 
distributions obtained with DPMJET-III-2011.
The experimental data for  {\em nsd} and  {\em inel} collisions are from the CMS ({\em nsd})
 \cite{cmsnsd} and from the ALICE \cite{alicensdinel} collaboration.}

The energy-dependence of the central density 
$dN/d\eta_{cm}$ at $\eta_{cm}$ = 0 
is presented in Fig.3  for p--p collisions of  {\em nsd} and  {\em inel} 
events. 
The DPMJET--III-2011 results are compared with data from various energies.
In all cases a good agreement is obtained.

\includegraphics[width=14cm]{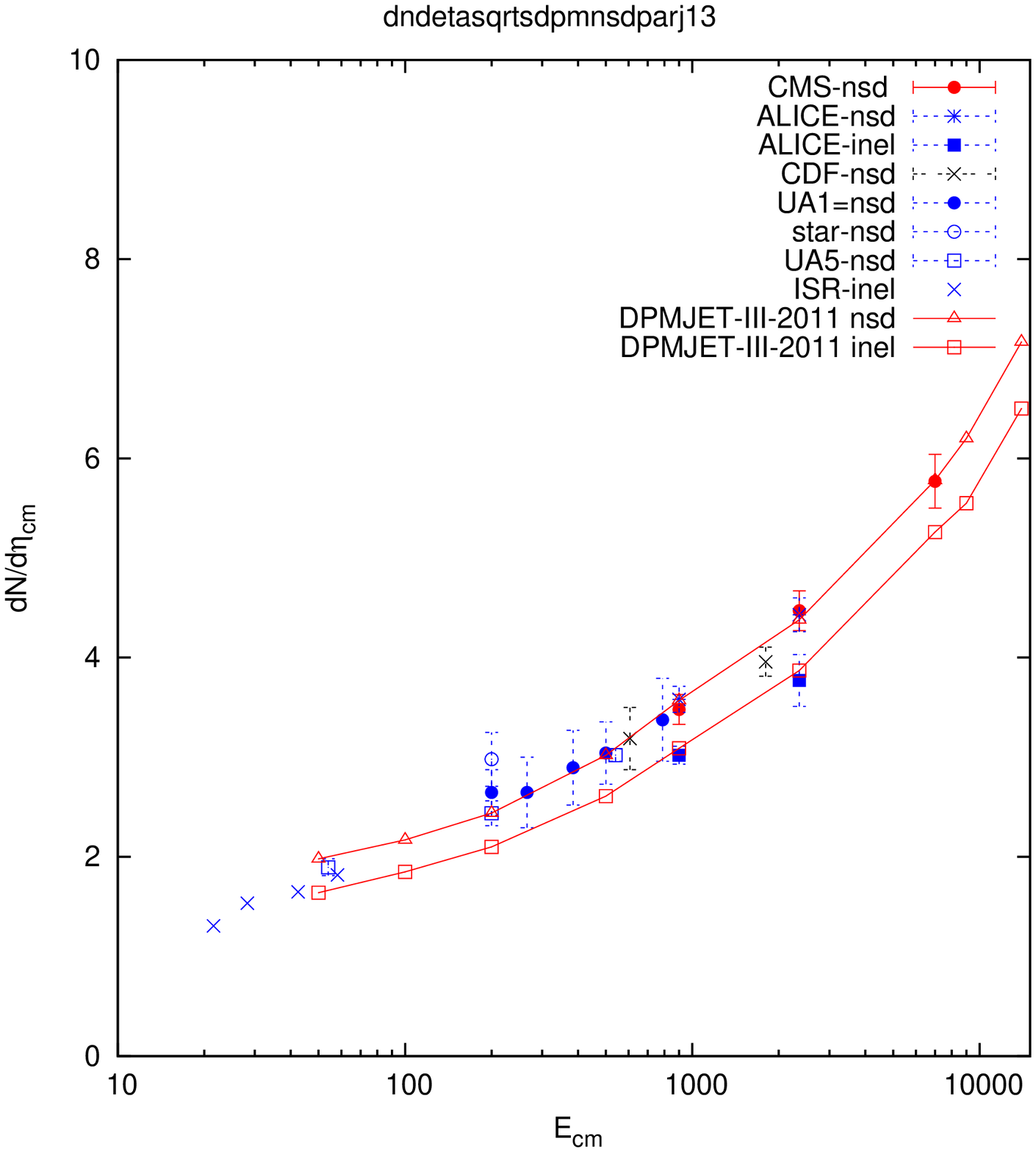} 

{{\bf Fig.3} Central $\eta_{cm}$ values of all charged particles 
in  p--p collisions compared to  DPMJET-III-2011 results.
The experimental  data are from the CMS 
collaboration~\cite{cmsnsd} for {\em nsd} collisions and from the ALICE
collaboration~\cite{alicensdinel} for {\em nsd} and  {\em inel} collisions. 
Further data are from UA5 \cite{ua5pbp}, the ISR \cite{isrinel},
 STAR \cite{starppnsd}, UA1 \cite{ua1nsdpbp} and CDF
 \cite{cdfnsdpbp}.      }

In Fig.4 we compare $p_t$ distributions from the 
DPMJET--III-2011  in p--p collisions at  $\sqrt s$ = 900, 2360 and
7000 GeV with representative experimental data points from the CMS
Collaboration \cite{cmsdpmjet}. 
The agreement between the modified
program  and the CMS data points is good.

\includegraphics[width=13cm]{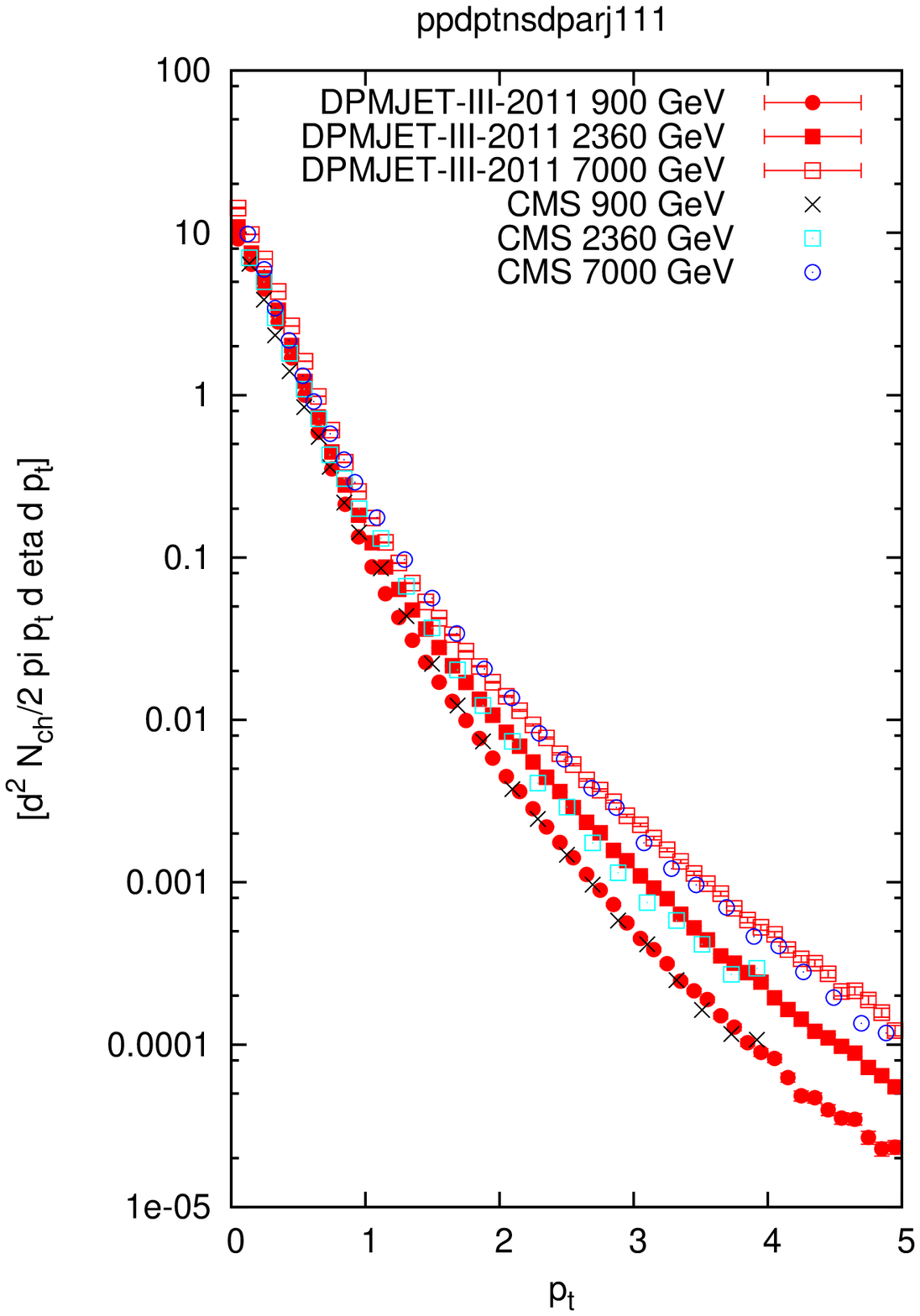} 

{{\bf Fig.4} Transverse momentum distributions in p--p
collisions at $\sqrt s$ = 900, 2360 and 7000 GeV. We compare
experimental data from the CMS Collaboration \cite{cmsdpmjet} to
DPMJET-III-2011 calculations.}

In Figs. 5-7 we compare the multiplicity distributions
for $|\eta| < 1. $ with experimental 
data from the ALICE Collaboration  \cite{alicensdinel}. 
Again  a reasonable agreement is obtained.

\includegraphics[width=8cm]{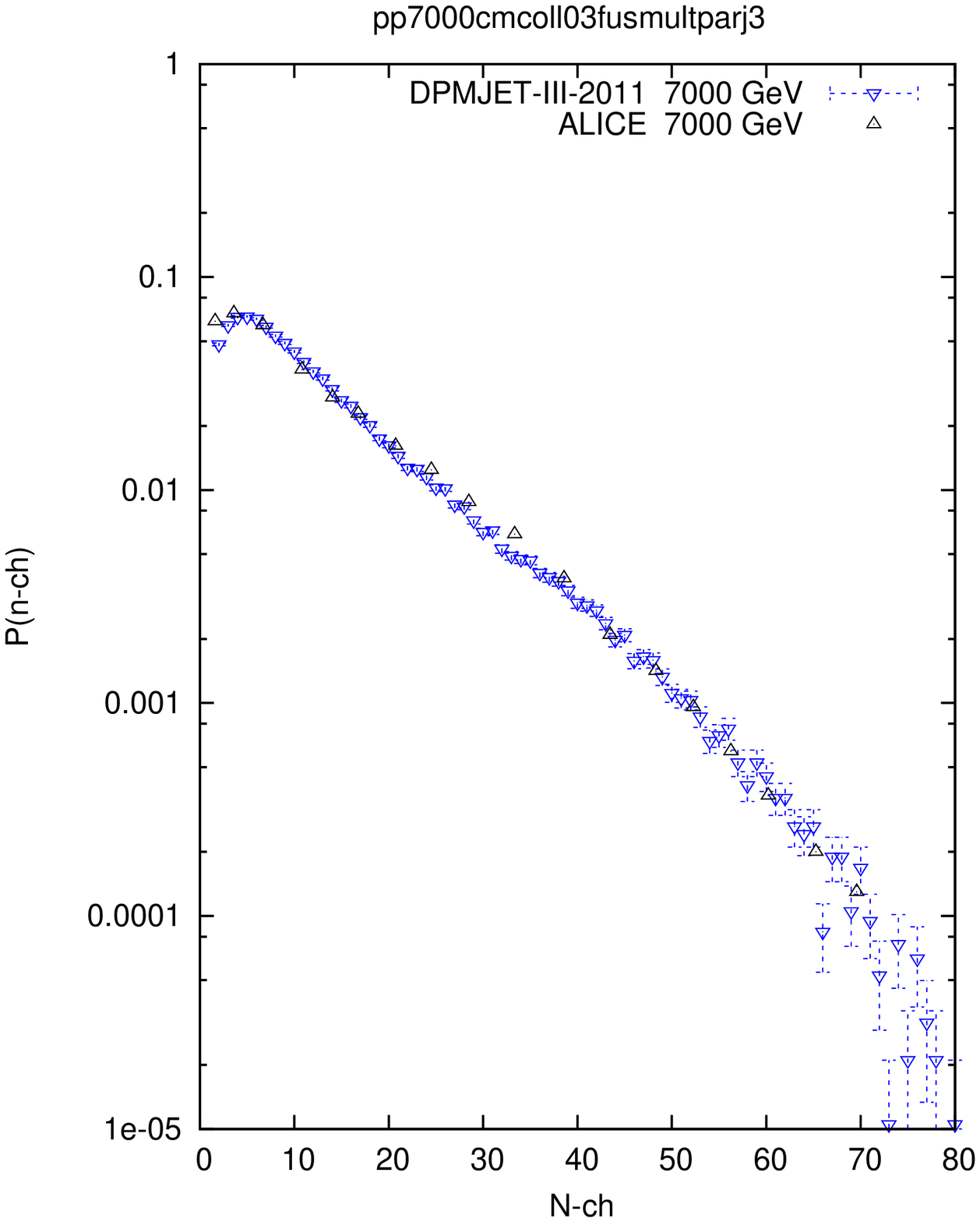} 

{{\bf Fig.5} Multiplicity distributions in p--p collisions at
$\sqrt s$ = 7000 GeV. We compare  data for $|\eta| <
1.$ from the ALICE Collaboration \cite{alicensdinel} to the
DPMJET-III-2011 result. }

\includegraphics[width=8cm]{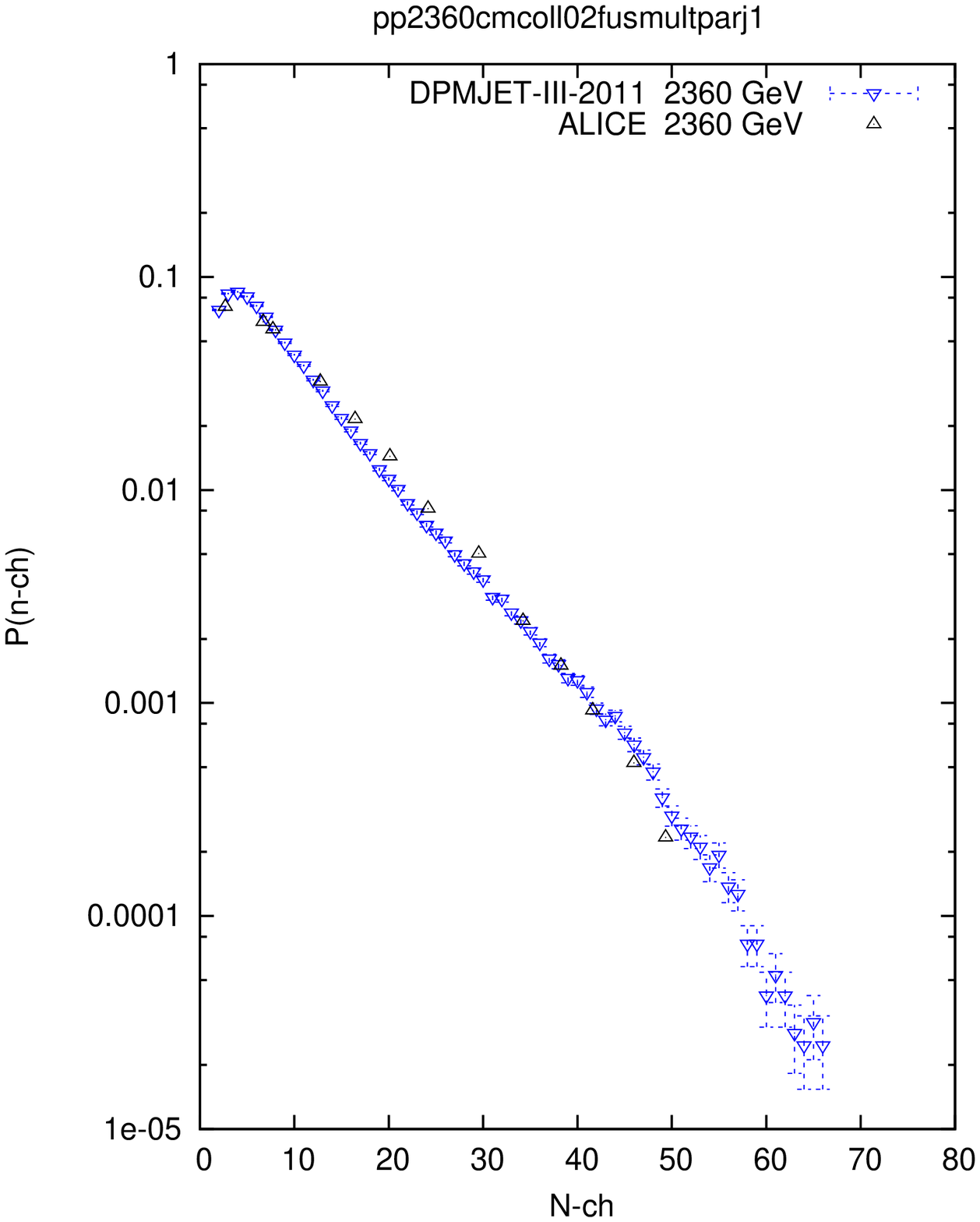} 

{{\bf Fig.6} Multiplicity distributions in p--p collisions at
$\sqrt s$ = 2360 GeV. We compare  data for $|\eta| <
1.$ from the ALICE Collaboration \cite{alicensdinel} to the
DPMJET-III-2011 result. }

\clearpage

\includegraphics[width=8cm]{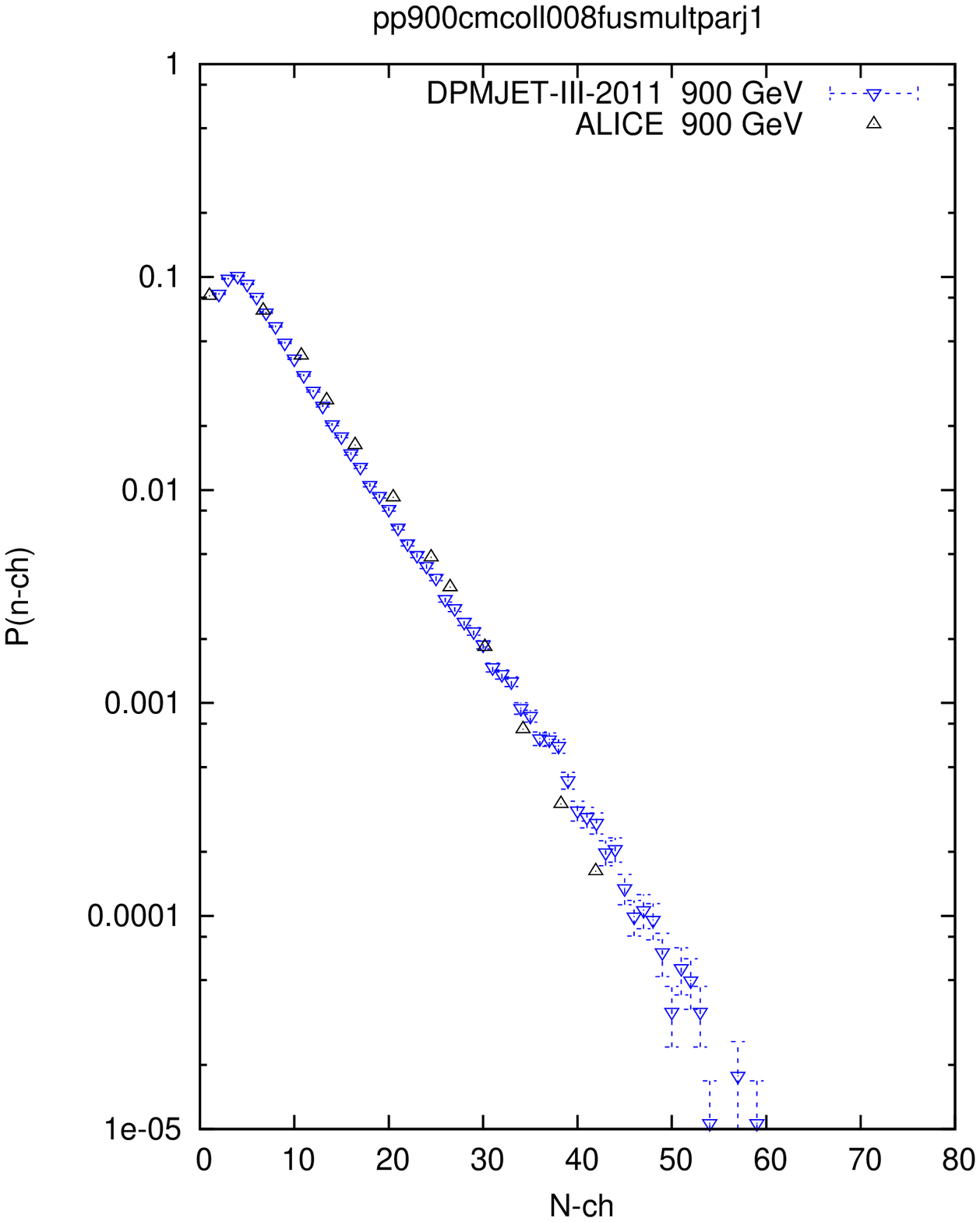} 

{{\bf Fig.7} Multiplicity distributions in p--p collisions at
$\sqrt s$ = 900 GeV. We compare data for $|\eta| <
1.$ from the ALICE Collaboration \cite{alicensdinel} to the 
DPMJET-III-2011 result. }

\section{Comparison of DPMJET--III-2011 results with LHC data on
the production of strange hadrons} 
 
 Strange hadron production in p-p
  collisions was measured by the CMS Collaboration \cite{CMSstrange}. It
determined the production of
 $K_s^0$
 mesons and the production of $\Lambda$ and $\Xi ^-$
 hyperons.

 Similar data on the production of strange hadrons were also given by
 the ALICE Corporation \cite{ALICEstrange}. We did not include them 
so far as they do not  affect the consideration discussed below.

 In Fig.8 we compare the production  of  $K_s^0$ and $\Lambda$
 hyperons  in dependence  of the $dn/dy_{cm}$ . 
 With  energy-dependent parameters good agreement of the  DPMJET-III-2011 and the CMS
 measurements is obtained. 
 
But the situation is not perfect. CMS also
 measures transverse momentum distributions. Comparing
 transverse momentum distributions we find the shape of the
 distributions to differ. This can be seen in Fig.9  showing the transverse
 momentum distributions of $\Lambda$ hyperons.
Above 1 GeV the model is below the data. A similar problem seems to appear in
many model calculations~\cite{Long}.

If we would also adjust the
 parameters in such a way, that the agreement between the
 transverse momentum distributions is optimal in this region,  we
 would obtain a disagreement in the  $dn/dy_{cm}$ distributions.

\includegraphics[width=15cm]{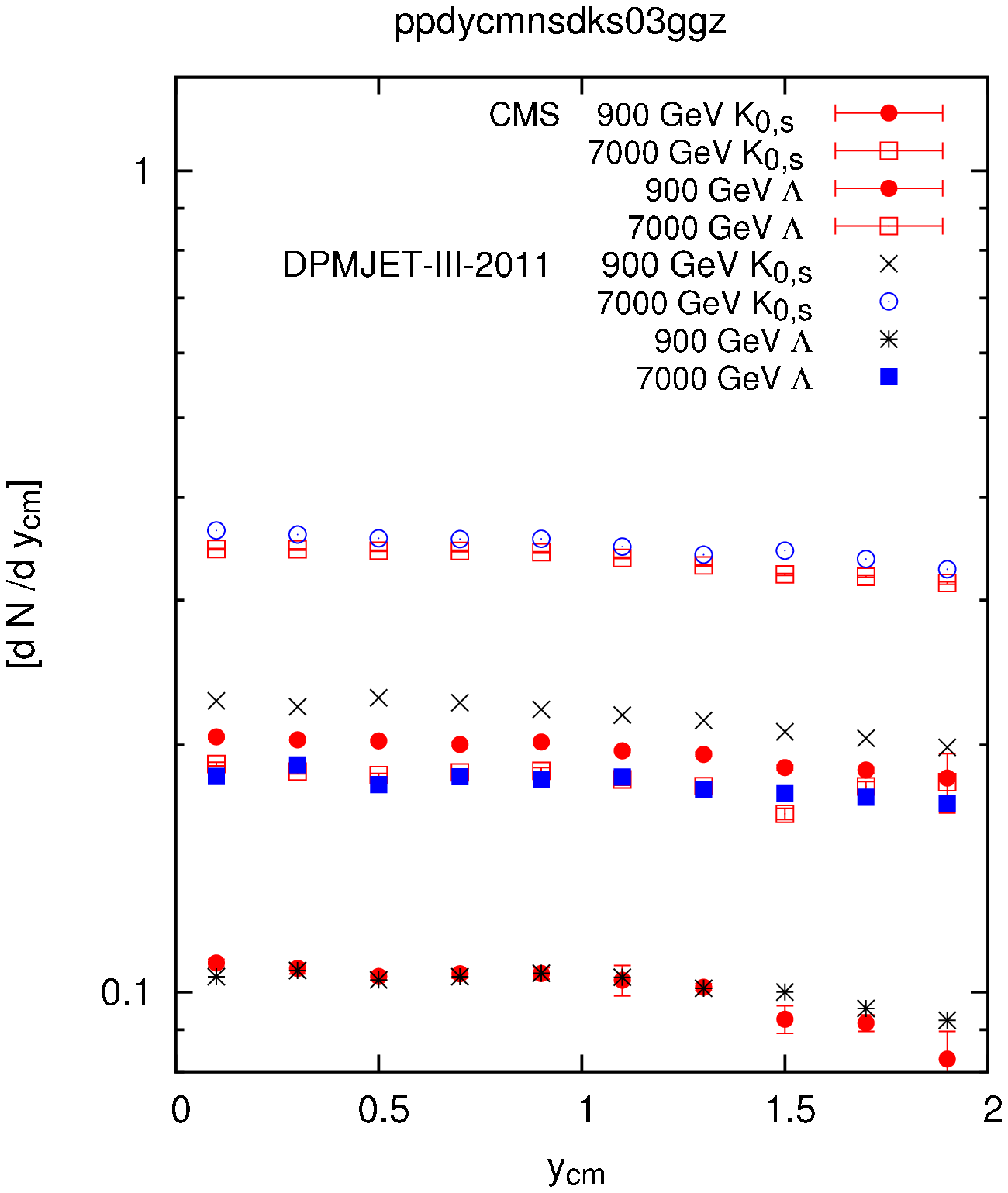} 

{{\bf Fig.8} $dn/dy_{cm}$ distributions in p-p collisions 
of $K_s^0$ and $\bar K_s^0$
as well as $\Lambda $ and $\bar\Lambda $. We compare LHC data 
 from the CMS Collaboration \cite{CMSstrange} to
DPMJET-III-2011 result. }

\newpage

\includegraphics[width=15cm]{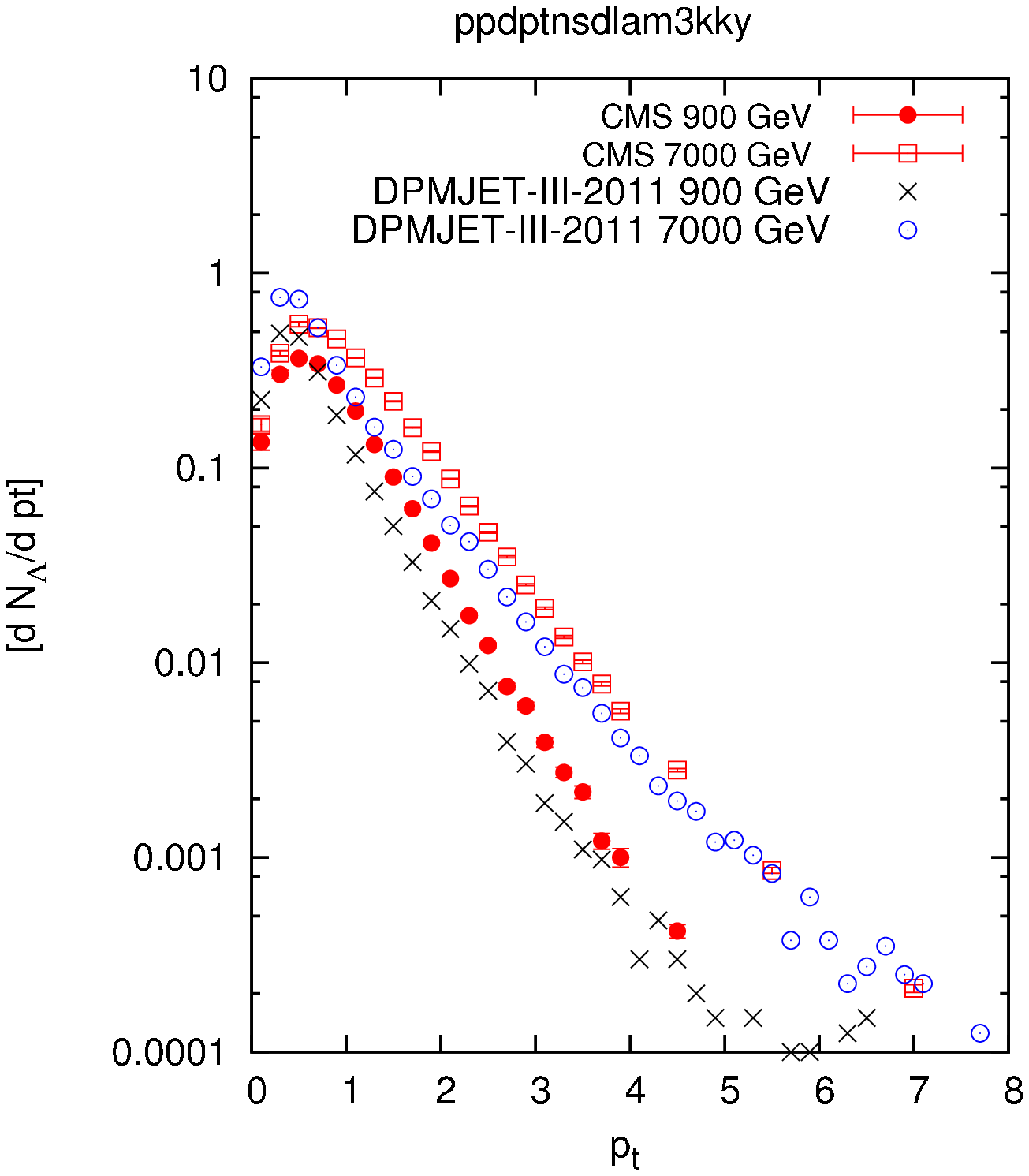} 

{{\bf Fig.9} Transverse momentum distributions of  
 $\Lambda$ and $\bar\Lambda$. We compare experimental data 
 from the CMS Collaboration \cite{CMSstrange} to
DPMJET-III-2011 result. }

 Unfortunately here the situation becomes even more  problematic.
 In Fig.10 we compare the  $dn/dy_{cm}$ distributions of  $\Xi$ hyperons in  the
 DPMJET-III-2011 with the measurements of CMS. The modified model 
 predicts  $\Xi$  distributions about three times as
 large as measured by CMS. We have modified the parameters in
 such a way, that the $\Lambda$ hyperons agree 
with the CMS data. The same parameters should also
 lead to agreement for the  $\Xi$  hyperons. They do not. We can
 only conclude, that so far we do not fully understand  the production
 of  $\Xi$ hyperons in DPMJET-III.

\includegraphics[width=12cm]{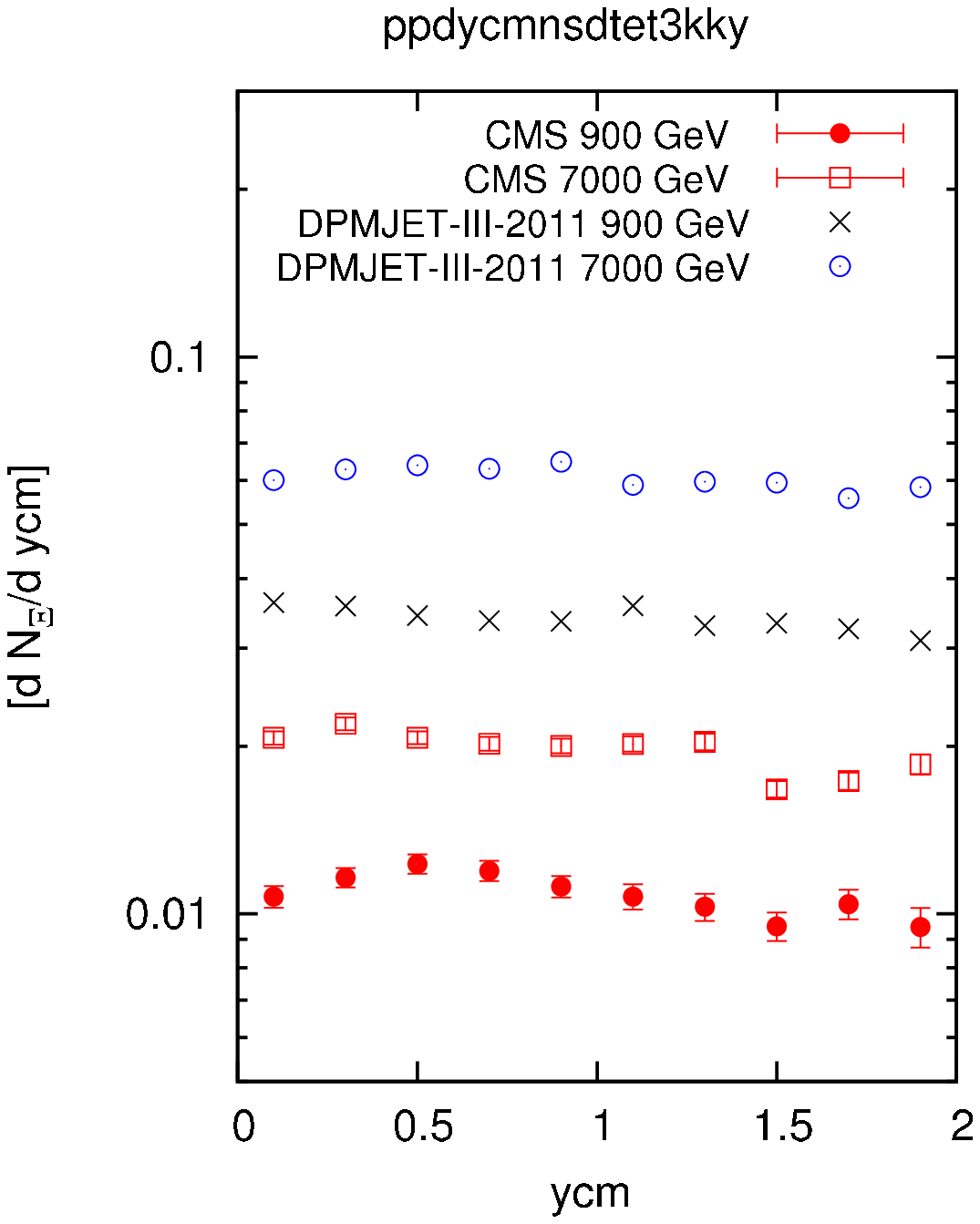} 

{{\bf Fig.10}  $dn/dy_{cm}$  distributions of  
 $\Xi$ and $\bar\Xi$. We compare experimental data 
 from the CMS Collaboration \cite{CMSstrange} to 
DPMJET-III-2011 calculations. }

\section{Conclusions} 
 
DPMJET-III is a code for hadron production in hadron-hadron,
photon-hadron, hadron-nucleus and nucleus-nucleus collisions
 \cite{phojet-a,dpmjet3} which is about 10 to 15 years old. The
 measurements of hadron production in p-p collisions at LHC
 energies gave the occasion to check how well DPMJET-III  agrees
 to the data in this  higher energy region. We knew already from
 comparisons at the lower FERMILAB energies, that not all
 features of DPMJET-III are valid at higher energies, a well
 known example is the collision scaling, which was found at
 FERMILAB in hadron-nucleus and nucleus-nucleus collisions and
 which is not a property of DPMJET-III.

 Comparing DPMJET-III with the LHC data we found further
 problems of DPMJET-III. The energy-dependence of hadron
 production measured by the LHC collaborations at 7 TeV did differ from
 the one predicted by the original DPMJET-III. In the present
 paper we find, that this energy-dependence can be corrected in
 DPMJET-III by making in DPMJET-III the PYTHIA-parameters energy
 dependent. We consider the introduction of energy-dependent
 parameters only as a temporary solution to get agreement of
 DPMJET-III with the new LHC data. 

A more  permanent solution will
require deeper changes in the program.
  Of course the parameters entering DPMJET-III, PHOJET and PYTHIA
  should not be adjusted per hand to determine the energy-dependence of the hadron
  production models. This energy-dependence should be an
  intrinsic property of the hadron production models.
  We conclude that we need a new version of 
  the model  which agrees better with
  the data in the new energy region opened by the LHC.


\section{Bibliography}



\begin{thebibliography}{99}

\bibitem{Aurenche:1994ev}
  P.~Aurenche, F.~W.~Bopp, R.~Engel, D.~Pertermann,
J.~Ranft, S.~Roesler,
 Comput.\ Phys.\ Commun.\  {\bf 83 } (1994)  107-123.
 [hep-ph/9402351]. 




\bibitem{phojet-a}
R.\ Engel:
Z. Phys. C \textbf{66}, 203 (1995),
R.\ Engel and J.\  Ranft:
Phys. Rev. D \textbf{54}, 4244 (1996)




\bibitem{dpmjet3}
S.\ Roesler, R.\ Engel and J.\  Ranft:
hep--ph/0012252, Proc. of Monte Carlo 2000, Lisboa, Oct.2000,
Springer,p.1033

 \bibitem{Sjostrand:2006za}
   T.~Sjostrand, S.~Mrenna, P.~Z.~Skands,
  JHEP {\bf 0605 } (2006)  026.
   [hep-ph/0603175]. 



\bibitem{dpmfusion1}
 J.\  Ranft, R.\ Engel and S.\ Roesler: Nucl.Phys. B
 (Proc.Suppl.) 122 (2003) 392

\bibitem{Bopp:2005cr}
   F.~W.~Bopp, J.~Ranft, R.~Engel, S.~Roesler,
   Phys.\ Rev.\  {\bf C77}, 014904 (2008).
   [hep-ph/0505035]. 




\bibitem{cmsdpmjet}
CMS Collaboration, hep.ex 1005.3299, hep.ex 1011.5531 and hep.ex
1012.1605.

\bibitem{alicedpmjet}
ALICE Collaboration, hep.ex 1004.3514, hep.ex 1007.0719, hep.ex
1010.2448 and hep.ex 1102.2369.

\bibitem{atlasdpmjet}
ATLAS Collaboration, hep.ex 1003.3124 and hep.ex 1010.0843.

\bibitem{collsca}
F.W.Bopp, J.Ranft, R.Engel and S.Roesler, hep-ph/0403084 based
on a poster subm. to the 17th Int.Conf.on Ultra relativistic
nucleus--nucleus collisions, Oakland, Calif. USA, 2004. RHIC
data and the multichain Monte Carlo DPMJET--III.

\bibitem{Klay:2004ma}
  J.~L.~Klay,
  J.\ Phys.\ G {\bf G31 } (2005)  S451-S464.
  [nucl-ex/0410033].

\bibitem{cmsnsd}
CMS Collaboration, Khachatryan V. et al., J. High Energy Phys.,
2010(2010) 02041.

\bibitem{alicensdinel}
ALICE Collaboration, Elia D., hep-ex 1102.2369.

\bibitem{ua5pbp}
UA5 Collaboration, Alner G. J. et al.,Z.Phys. C, 33 (1986) 1.

\bibitem{isrinel}
Aachen--CERN--Heidelberg--Munich Collaboration, Nucl.Phys. B129
(1977) 365.

\bibitem{starppnsd}
STAR Collaboration, Phys.Rev. C79 (2009) 034909.

\bibitem{ua1nsdpbp}
UA1 Collaboration, Nucl. Phys. B335 (1990) 261.

\bibitem{cdfnsdpbp}
CDF Collaboration, Phys. Rev. D41 (1990) 2330.


\bibitem{CMSstrange}
CMS Collaboration, hep-ex 1102.4282 (2011) and with corrections
as given in http://hepdata.cedar.ac.uk/view/p8016.

\bibitem{ALICEstrange}
K.~Aamodt, A.~Abrahantes Quintana, D.~Adamova, A.~M.~Adare,
M.~M.~Aggarwal, G.~Aglieri Rinella, A.~G.~Agocs, S.~Aguilar Salazar {\it et
al.},
  Eur.\ Phys.\ J.\  {\bf C71 } (2011)  1594.
  [arXiv:1012.3257 [hep-ex]].

\bibitem{Long}
Hai-Yan Long et al., hep-ph 1103.2618, 2011.



\end{thebibliography}
\end{document}